\documentclass{article}

\usepackage{latexsym}
\newtheorem{theorem}{Theorem}
\newtheorem{lemma}{Lemma}
\newtheorem{corollary}{Corollary}
\newtheorem{fact}[lemma]{Fact}

\newtheorem{proposition}{Proposition}

\newcommand{\trace}{{\rm Tr}}

\newcommand{\ket}[1]{\left|#1\right\rangle}
\newcommand{\bra}[1]{\left\langle #1\right|}

\newcommand{\ketbra}[2]{\ket{#1}\!\bra{#2}}
\newcommand{\norm}[1]{\left\|\,#1\,\right\|}
\newcommand{\trnorm}[1]{\norm{#1}_{\rm t}}

\newcommand{\co}{{\cal O}}
\newcommand{\ch}{{\cal H}}

\newcommand{\brho}{{\mathbf \rho}}

\newcommand{\trn}[1]{\trnorm{#1}}

\newenvironment{proof}[1]{%
\begin{trivlist}{}{\setlength{\topsep}{0cm}\setlength{\partopsep}{0cm}}
\item \textbf{#1.\@}\hspace*{1ex}\ignorespaces}%
{\makebox[0cm]{}\nolinebreak\hfill$\Box$\end{trivlist}}

\date{}
\begin{document}

\title{Quantum Time-Space Tradeoffs for Sorting
\footnote{An extended abstract of this paper appears in the proceedings of STOC'03.} }

\author{Hartmut Klauck\thanks {Supported by NSF grant
CCR-9987845.}\\
      School of Mathematics\\
       Institute for Advanced Study\\
       Princeton, NJ08540, USA\\
       {\tt klauck@ias.edu}
} \maketitle

\begin{abstract}
We investigate the complexity of sorting in the model of sequential
quantum circuits. While it is known that in general a quantum algorithm
based on comparisons alone cannot outperform classical sorting algorithms by
more than a constant factor in time complexity, this is wrong in a space
bounded setting. We observe that for all storage bounds $n/\log n\ge S\ge
\log^3 n$, one can devise a quantum algorithm that sorts $n$ numbers
(using comparisons only) in time $T=O(n^{3/2}\log^{3/2} n/\sqrt S)$.
We then show the following lower bound on the time-space tradeoff for
sorting $n$ numbers from a polynomial size range in a general sorting
algorithm (not necessarily based on comparisons): $TS=\Omega(n^{3/2})$.
Hence for small values of $S$ the upper bound is almost tight.
Classically the time-space tradeoff for sorting is $TS=\Theta(n^2)$.
\end{abstract}

\section{Introduction}
\label{sec-intro}

Sorting is arguably one of the most important and well-studied problems
in computer science. While any compari\-son-based algorithm needs
$\Omega(n\log n)$ time to sort $n$ numbers, and several algorithms
achieving a matching running time are well known, the situation changes if we
are confronted with the problem to sort when the list of data is too long to fit
into the memory of our computer. In this setting we are given a bound
$S(n)$ on the available memory and have to sort with an algorithm whose
space requirements do not exceed this bound. This situation arises
e.g.~if data are stored distributed on the internet and are too
large to be held in the memory of the computer at a single site.

After prior results by Borodin and Cook in \cite{BC82}, Beame \cite{B91}
proved a lower bound of $TS=\Omega(n^2)$ for any classical algorithm
sorting $n$ numbers. A matching upper bound for sorting on a RAM has
been proved in \cite{PR98}, for all space bounds $\log n\le S\le n/\log
n$.
This upper bound is actually achieved by an algorithm that accesses the
data only using comparisons. This is remarkable, since sometimes
operations on the numbers itself can speedup sorting, e.g.~Radixsort
(see e.g.~\cite{CLRS01}) can sort polynomial size numbers in linear time,
while comparison based algorithms need time $\Omega(n\log n)$ for this
task. So the classical time-space complexity of sorting is well known.
Interestingly, while Beame \cite{B91} investigates the {\sc Unique
Elements} problem (output the list of elements appearing exactly once in
the input), Borodin and Cook \cite{BC82} consider the {\sc Ranking} problem
(output the permutation needed to sort) and show  $TS=\Omega(n^2/\log
n)$ for numbers from a quadratic size range.

Quantum computing is an active research area offering interesting
possibilities to obtain improved solutions to information processing
tasks by
employing computing devices based on quantum physics, see
e.g.~\cite{NC00} for a nice introduction into the field. It has been
shown by H\o yer et al.~in \cite{HNS01}, however, that any comparison-based quantum
sorting algorithm needs time $\Omega(n\log n)$, which is already
achieved by
classical algorithms. So is there no point in using quantum computers to
sort? We demonstrate that in the space bounded setting the quantum
complexity of sorting is quite different from the classical complexity.

We use the model of quantum circuits to investigate time-space tradeoffs.
While in the classical setting branching programs are the standard model
to consider these problems, employing quantum branching programs seems
to complicate the definition of our model unnecessarily. A quantum
circuit uses space $S$, if it operates on $S$ working qubits. Furthermore it has
write-only access to several output qubits, since this output is
typically too large to be held in the working memory during the whole
computation. The circuit accesses the input by specific oracle gates. For
the upper bound we use a comparison oracle, i.e., an oracle gate gets
(superpositions over) two indices of numbers to be compared and outputs a
bit indicating the comparison result into an extra qubit. For the lower
bound we choose to restrict the size of the numbers in the input to
$n^2$, and here we consider an access oracle, that directly reads
numbers from the input into the work space.

Using Grover's famous quantum search algorithm and its adaptation to minimum
finding by D\"urr and H\o yer in \cite{DH96} we can show the following:

\begin{theorem}\label{the:alg}
For all $S$ in $[\Omega(\log^3 n),\ldots, O(n/\log n)]$ there is a
quantum circuit with space $S$ that, given a comparison oracle for $n$
numbers, outputs the sorted sequence, and uses time $O(n^{3/2}\log^{3/2}
n/\sqrt S)$. The entire output is correct with probability $1-\epsilon$
for an arbitrarily small constant $\epsilon>0$.
\end{theorem}

If we restrict the size of the numbers, we can use the same algorithm
while employing an access oracle.

\begin{corollary}\label{cor:alg}
For all $S$ in $[\Omega(\log^3 n),\ldots, O(n/\log n)]$ there is a quantum circuit with space $S$ that, given an access oracle for $n$ numbers from
$\{1,\ldots,n^k\}$ for some constant $k$, outputs the sorted sequence in time $O(n^{3/2}\log^{3/2} n/\sqrt S)$. The entire output is correct with
probability $1-\epsilon$ for an arbitrarily small constant $\epsilon>0$.
\end{corollary}

The quantum sorting algorithm clearly beats the classical time-space
tradeoff lower bound. Actually it gives rise to a tradeoff
$T^2S=O(n^3\log^3 n)$. It performs best in comparison to classical algorithms whenever $S$ is small. E.g.~for $S=\log^k(n)$
the running time is $O(n^{3/2}/\log^{(k-3)/2}n)$, a clear improvement
compared to the classical bound $\Omega(n^2/\log^k n)$.

We then investigate the question whether one can do even better. First
let us fix an output convention for sorting. Denote by $Min(x,i)$ the $i$th number
of the sorted sequence.
Let us assume an algorithm outputs a sequence $(y_1,j_1),\ldots, (y_n,j_n)$, so that the set
$\{j_1,\ldots,j_n\}$ equals $\{1,\ldots,n\}$. Since we consider quantum
algorithms we will allow errors. Furthermore we will require that the
$j_i$ are produced in the same order for all inputs.
Our main result is the following lower bound.

\begin{theorem}\label{the:lower}
Let $A$ be any quantum circuit that, given an access oracle for a
sequence $x$ of $n$ numbers from $\{1,\ldots, n^2\}$, outputs $n$ pairs $O_1(x),\ldots, O_n(x)$, so that the probability that
$O_i(x)=(Min(x,j_i),j_i)$ is $2/3$ for each $i$ and the $j_i$ are a fixed permutation of $\{1,\ldots,n\}$.
Suppose $A$ uses $S$ work qubits and $T$ oracle gates, then
$ST=\Omega(n^{3/2})$.
\end{theorem}

Note that we do not even require that the whole output sequence is ever
simultaneously correct.

Hence for small $S$ we cannot substantially speed up the quantum sorting
algorithm. E.g.~for $S=poly(\log n)$ the necessary and sufficient time
to sort on a quantum computer is $\widetilde{\Theta}(n^{3/2})$.

Technically the lower bound proof proceeds as follows. A given quantum
circuit is sliced into parts containing only $\delta\sqrt n$ queries.
Each such slice starts with some initial information stored in the $S$ qubits
that has been computed by the previous slices. We show how to get rid of
this initial information while deteriorating the success probability
only exponentially in $S$. This step is similar to an application of the
union bound (classically the success probability for one of the $2^S$
fixed states of the $S$ bits given in the beginning must be $1/2^S$ times
as good as the success probability with initial information). We show
that something similar is possible in the quantum case.

Then we are left with the problem to analyze the success probability of
a slice of the circuit without initial information, and show that it can
be only large enough, if the number of outputs is small. Since there
have to be $n$ outputs this leads to a lower bound on the number of
slices and hence the time complexity.

We also obtain a tight analysis for classical sorting in the case that all the numbers are known to be distinct.
No tight analysis of {\sc Ranking} seems to have appeared since \cite{BC82} (only a modest improvement in \cite{RS82}),
and the {\sc Unique Elements} problem is
not fit to provide lower bounds in the situation when
all the inputs are known to be distinct, which they are with high
probability when drawn uniformly from a large range.

\section{Definitions and Preliminaries}\label{sec-def}

In this section we give some background on quantum states and their
distinguishability, define the model of quantum circuits we are studying,
describe a quantum version of the union bound, and discuss lower bounds
on query complexity by query magnitude arguments. For more quantum
background see \cite{NC00}.

\subsection{Quantum States}

The quantum mechanical analogue of a random variable is a probability
distribution over superpositions, also called a {\em mixed state}. For
the mixed state~$X = \{p_i,\ket{\phi_i}\}$, where $\ket{\phi_i}$ has
probability~$p_i$, the {\em density matrix\/} is defined as~$\brho_X
= \sum_i p_i \ketbra{\phi_i}{\phi_i}$. Density matrices are Hermitian, positive
semi\-definite, and have trace $1$. I.e., a density matrix has
real eigenvalues between zero and
one, and they sum up to one.

The {\em trace norm\/} of a matrix~$A$ is defined as~$\trn{A} =
\trace\,{\sqrt{A^\dagger A}}$, which is the sum of the magnitudes of the
singular values of~$A$. Note that if $\rho$ is a density matrix, then it has
trace norm one.

A useful theorem states that for two mixed states $\rho_1,\rho_2$ their
distinguishability is reflected in $\trn{\rho_1-\rho_2}$ \cite{AKN98}:

\begin{fact}
\label{fact:trace} Let~$\rho_1,\rho_2$ be two density matrices on the
same space~$\ch$. Then for any measurement~$\co$,

$$
\norm{\rho_1^\co - \rho_2^\co}_1 ~~\le~~ \trn{\rho_1-\rho_2},
$$

where~$\rho^\co$ denotes the classical distribution on outcomes
resulting from the measurement of~$\rho$, and~$\norm{\cdot}_1$ is
the~$\ell_1$ norm. Furthermore, there is a measurement $\co$, for which the above is
an equality.
\end{fact}

\subsection{Quantum Circuits}

We now define quantum oracle circuits. The two parameters we are
interested in are the number of work qubits corresponding to the space
bound, and the number of queries, which is always smaller than the overall
number of gates, corresponding to the time.

A {\it quantum circuit} on $S$ work qubits and $M$ output qubits is
defined as an ordered set of gates, where each gate consists of a unitary
operation on some $k$ qubits and a specification of these $k$ qubits.
The gates either operate on the work qubits only, or they are output gates that perform a
controlled-not, where the control is among the work qubits, and the
target qubit is
an output qubit. We require that each output qubit is used only once. In
this way the output qubits are usable only to record the computation
results.

Another type of gates are {\it query gates}, and they are needed to
access the input. In general a query gate performs the following
operation for some input $x\in\{1,\ldots,2^k\}^n$:
\[|i\rangle|a\rangle\mapsto|i\rangle|a\oplus x_i\rangle,\] for all
classical strings $i,a$, where the length
of $i$ is $\log n$ and the length of $a$ is $k$. The behavior of a query
gate on superpositions is defined by linearity.
In a {\it computation} all qubits start blank, and then the gates are
applied in the order defined on them.
For sorting we consider two different types of input oracles. In a {\it
comparison} oracle the algorithm is allowed to query an $n^2$-bit string
that contains the results of the pairwise comparisons between $n$
numbers. In an {\it access} oracle to input $x_1,\ldots,x_n$ the query
gate directly reads numbers $x_i$ (or their superpositions).
Note that an access oracle can efficiently simulate a comparison oracle if the numbers to be sorted are from a polynomial size range.

We require that a sorting circuit makes outputs in the following way: if
the input is presented as a comparison oracle, then the algorithm is
supposed to output the inverse of the permutation needed to sort. If the
input is given as an access oracle, then the output is given as a
list of numbers with their positions in the sorted sequence. We require that a number is output as a whole at
some point in the algorithm, i.e., there are $k$ one bit output gates
directly following each other when a number from a $2^k$ range is output.

Let us consider the output convention more closely (in the case of an
access oracle). For some input $x$ let $O_i$ denote the $i$th output, and
$O_i(x)$ the probability distribution obtained by measuring this output.
This is a distribution on pairs $(y_i,j_i)$.
Let $Min(x,i)$ denote the $i$th smallest element of the sorted sequence in the
input, then the pair $(Min(x,j_i),j_i)$ is supposed to be produced at output $O_i$.
The sequence $(j_i)$ is independent of $x$.
We say that $O_i(x)$ is correct with probability $1-\epsilon$, if
$O_i(x)=(Min(x,j_i),j_i)$ with probability $1-\epsilon$. We require in our
lower bound theorem that each $O_i(x)$ is correct with probability 2/3.

A circuit is said to have {\it time complexity} $T$, if the number of
gates is $T$ (in the lower bound we will only count the number of queries).

\subsection{A Quantum Union Bound}

In this subsection we want to develop a simple tool needed to take away
the initial information in a slice of the sorting circuit.

Let $\rho$ denote the density matrix of a state on $m$ qubits. We want
to replace $\rho$ by the completely mixed state, while retaining some of
the success probability of an algorithm taking $\rho$ as an input. The
following lemma will be helpful.

\begin{lemma}\label{lem:overlap}
Let $\rho$ be any density matrix on $m$ qubits. Let $M$ denote the
density matrix of the completely mixed state, i.e., the matrix with
entries
$1/2^m$ on the diagonal and zeros elsewhere. Then there exists a density
matrix $\sigma$, so that
\[M=1/2^m \rho + (1-1/2^m)\sigma.\]
\end{lemma}

\begin{proof}{Proof} We have to show that $\sigma=(M-\rho/2^m)/(1-1/2^m)$ is
a density matrix. For $\rho$ being a density matrix, $\sigma$ is clearly
Hermitian and has trace 1. So we have to show that $\sigma$ is positive
semidefinite, which is equivalent to $I-\rho$ being positive
semidefinite, for the identity matrix $I$. Let $U$ be some unitary
transformation that diagonalizes $\rho$, i.e., $D=U\rho U^\dagger$ is
diagonal. $U$ exists, since $\rho$ is Hermitian. Clearly $I-\rho$ is
positive semidefinite iff $I-D$ is, and $D$ contains on its diagonal
nonnegative numbers that sum to 1, hence $I-D\ge 0$.\end{proof}

We want to apply the above lemma in the following way.

\begin{lemma}\label{lem:union}
Suppose there is an algorithm that on some input $x$ first receives $S$
qubits of initial information depending arbitrarily on $x$, and that
makes afterwards only queries to the input. Suppose the algorithm
produces some output correctly with probability $p$.

Then there is an algorithm that uses no initial information, makes the
same number of queries, and has success probability $p/2^S$.
\end{lemma}

Actually the above lemma can be thought of as a quantum union bound.
Note that the $S$ qubits can be in as many states as there are inputs
$x$, still removing them decreases the success probability by a factor
exponentially in $S$ only.

\begin{proof}{Proof} Replace the quantum state containing the initial
information by the completely mixed state $M$ on $S$ qubits. Then run
the algorithm
in exactly the same manner as before. Clearly the algorithm does not get
initial information in this way. Due to Lemma~\ref{lem:overlap} the
original state has some probability $1/2^S$ in $M$, respectively one can
view $M$ as a mixture of the original state with probability $1/2^S$ and
another state with the remaining probability. So also the outcome of the
algorithm is such a mixture, and if the success probability was $p$
originally, it must be at least $p/2^S$ for the modified algorithm.
\end{proof}

Note that the completely mixed state on $S$ qubits can easily be obtained from
a blank state on $2S$ qubits by performing Hadamard gates on the
qubits $1$ to $S$ and then controlled-not gates on the pairs $i,S+i$.

Another way of obtaining a similar result is to use quantum
teleportation (invented by \cite{BBC+93}, see also \cite{NC00}). In the
teleportation scheme two players who share $m$ EPR-pairs can communicate an arbitrary
quantum state in the following way. If player Alice holds a quantum state
$\rho$ on $m$ qubits, she applies measurements in the Bell basis to the
$m$ pairs of qubits given by one qubit from an EPR-pair and one qubit of
$\rho$ each. Bob, holding the $m$ other qubits belonging to the
EPR-pairs then gets a message from Alice containing her measurement
results in $2m$ classical bits. He is then able to perform certain operations on
his qubits depending on the message, which enable him to recover $\rho$.

It is known that for each $\rho$, the probability of each of the
possible measurement results is exactly $4^{-m}$. Furthermore for one of
the measurement results, Bob does not have to do anything to his qubits to
get $\rho$, i.e., with probability $4^{-m}$ the state of Bob is correct
already. Note that this implies that Bob's qubits before he receives the
message, which are in a completely mixed state, can be viewed as an
ensemble of states $\rho_1,\ldots, \rho_{4^m}$, where $M=\sum_i
4^{-m}\rho_i$, and $\rho_1=\rho$.

\subsection{Lower Bounds for Query Algorithms}

In \cite{BBBV97} an $\Omega(\sqrt{n})$ lower bound for the problem of
finding a marked element in an unordered database of size $n$ has been
given, matching the upper bound of Grover's algorithm \cite{G96}. This
lower bound relies on the notion of {\it query magnitude}. For other
lower bound techniques for query complexity see e.g.~\cite{BW02}.

The query magnitude technique is basically an adversary argument. An
adversary is able to change the black-box input without the query
algorithm noticing that (for a more refined type of quantum adversary argument see \cite{A00}).
We use the following  statement derived as Corollary
3.4 in \cite{BBBV97}.

\begin{fact}\label{fac:qm}
Let $x=x_1,\ldots,x_n$ be an input given as an access oracle, with
$x_i$ from some finite set, and let $x'(i)$ be any input that differs
from $x$ in position $i$ and nowhere else. $A$ is any quantum algorithm that
accesses the oracle via at most $T$ queries. The state $\rho_x$ denotes
the final state of $A$'s workspace when querying $x$, the state $\rho_{x'(i)}$
when querying $x'(i)$.

Then for any $\alpha>0$ there is a set of at least $n-T^2/\alpha^2$
input positions $i$ such that for all $x'(i)$:
$\trnorm{\rho_x-\rho_{x'(i)}}\le 2\alpha$.
\end{fact}

In our lower bound we will need a somewhat stronger statement that allows us to deal with a situation
conditioned on results of previous measurements.

\begin{lemma}\label{lem:condadv}
Let $x=x_1,\ldots,x_n$ be an input given as an access oracle, with
$x_i$ from some finite set, and let $x'(i)$ be any input that differs
from $x$ in position $i$ and nowhere else.
$A$ is any quantum algorithm that
accesses the oracle via at most $T$ queries. Suppose $A$ contains no measurements, but at the end a measurement is
performed on some of the qubits. Fix some outcome $F$ of this measurement that occurs with probability $q_x$. Assume
that some event $E$ happens with probability $p_x$ conditioned on $F$.

Then for any $\alpha>0$ there is a set of at least $n-T^2/\alpha^2$
input positions $i$ such that if $A$ is performed on $x'(i)$,
then the probability that $F$ is the outcome of the measurement and $E$ happens is at least $q_x(p_x-\alpha)$.
\end{lemma}

\begin{proof}{Proof}
Let $U_x$ and $U_{x'(i)}$ denote the unitary transformations done
by the circuit $A$ on inputs $x$ and $x'(i)$. W.l.o.g.~$A$ starts from the blank state $|0\rangle$ on
some qubits. Let $|\phi_x\rangle$ denote the state obtained by performing $U_x|0\rangle$, and measuring
the resulting state, when $F$ is the result of the measurement. Let $|\psi_x\rangle=U^{-1}_x |\phi_x\rangle$.
Clearly there is a state $|\theta_x\rangle$ so that $|0\rangle=\gamma_1 |\psi_x\rangle+\gamma_2|\theta_x\rangle$,
and $|\gamma_1|^2=q_x$, and $\langle \theta_x|\psi_x\rangle=0$. We can apply Fact~\ref{fac:qm} to $A$ running on state
$|\psi_x\rangle$, and get the required number of $x'(i)$, so that the obtained states are close to each other, i.e., $E$
happens with probability $p_x-\alpha$ when $U_{x'(i)}$ is applied to $|\psi_x\rangle$. We are interested in the joint
probability of events $E,F$
when $U_{x'(i)}$ is performed on $|0\rangle=\gamma_1 |\psi_x\rangle+\gamma_2|\theta_x\rangle$. Clearly this probability
is at least $q_x(p_x-\alpha)$.
\end{proof}

\section{A Sorting Algorithm}\label{sec-alg}

In this section we describe the algorithm needed to prove
Theorem~\ref{the:alg}. For this upper bound we identify time complexity with the
number of constant fan-in gates needed to build the circuit (queries still count as one gate).

The algorithm iterates minimum finding and uses
the following result described by D\"urr and H\o yer \cite{DH96} based on Grover's
famous search algorithm \cite{G96}.

\begin{fact}\label{fac:min}
There is a quantum query algorithm that, given a comparison oracle to
$n$ numbers, finds the minimum of these numbers with probability
$1-\epsilon$, and uses $O(\sqrt{n}\log(1/\epsilon))$ queries (and gates) and space $O(\log^2 n\log(1/\epsilon))$.
\end{fact}

Let $S$ be the space bound. Then $x=x_1,\ldots,x_n$ can be partitioned
into $b=S/(c\log n)$ blocks $y^i$ of $O(n\log n/S)$ numbers
each (for some large enough constant $c$).
Here is the algorithm:

\begin{enumerate}
\item FOR $i:=1$ to $b$ compute the position of the minimum of $y^i$
and store it together with $i$.
\item Arrange these minima positions as
a Heap ordered by the minima's size (see \cite{CLRS01}).
\item For $i:=1$ to $n$ DO
\begin{enumerate}
\item Output $(i,j)$ if the minimal number among the block minima is
$x_j$.
\item Remove $j$ and its block number $k$ from the Heap.
\item Find the position of the minimal number $x_l$ larger than
$x_j$ in $y^k$.
\item Insert $(l,k)$ into the Heap.
\end{enumerate}
\end{enumerate}

Note that all these operations can be performed with a comparison oracle.

Steps 1.~and 3.c) employ the minimum finding algorithm of
Fact~\ref{fac:min}. To ensure correctness in step 1.~that algorithm is
used with error bound $1/S^2$, hence the time for step 1.~is $O(S/\log(n)\cdot
\sqrt{n\cdot\log(n)/S}\cdot\log S)=O(\sqrt{n\log(n)S})$.

Step 2.~can be done in time $O(S)$, steps 3.a) and 3.b) need time $O(\log n)$,
and step 3.d) needs time $O(\log S\log n)$ in each iteration \cite{CLRS01}.

In step 3.c) the algorithm is used with error $1/n^2$. Then the running
time for 3.c) is overall $O(n\cdot \sqrt{n\cdot\log(n)/S}\cdot\log
n)=O(n^{3/2}\log^{3/2}(n)/\sqrt S)$.

The time spent in the other steps is dominated by the
time used in step 3.c), if $S=O(n/\log n)$.

Note that for reusing the $O(\log^3 n)$ qubits of storage needed by the
minimum finding algorithm in each iteration, it is understood that this algorithm is used
in the following way. It is run in the usual way, with measurements
deferred to the end. Then (instead of measuring) its output is copied to
some qubits using controlled-nots. Afterwards the minimum finding algorithm
is run {\it backwards} to clean up the storage it has used. Since the
error of the algorithm is small enough this leads to an algorithm with
overall error bounded by $O(1/n)$, compare \cite{BBBV97,AKN98} for
details on how to run subroutines on a quantum computer.

So overall the storage bound is not violated. Hence the algorithm
behaves as announced.

\section{The Lower Bound}\label{sec-lower}

Now we give the proof of Theorem~\ref{the:lower}. After some simple
preparations we show how to decompose a quantum circuit into slices that
contain only few oracle gates but must (on average) produce many
outputs. Then, in our main lemma, we give an upper bound on the number of
outputs such a slice can give. In the rest of this section we prove that
lemma.

\subsection{Preparations}

Let $A$ be a quantum circuit with $T$ oracle gates and $S$ work qubits.
$A$ contains $n$ output operations, each of which writes on $3\log n$ of
the output qubits. If one of these outputs is measured in the
standard basis, then with probability $2/3$ a pair $(Min(x,j_i),j_i)$ is produced,
and the $j_i$ form a fixed permutation of $\{1,\ldots,n\}$.

First we note that the success probability can be improved in some sense.

\begin{lemma}\label{lem:boost}
The success probability can be improved to $1-\epsilon$ for any constant
$\epsilon>0$ without changing $S,T$ by more than a constant factor, at the expense of
adding a circuit consisting of $O(\log n)$ qubits and a majority computation to any output gate.
\end{lemma}

\begin{proof}{Proof} We can use the circuit some $l$ times ``in parallel''. For
each output first all $l$ ``parallel'' outputs are mapped to some extra
work storage ($O(l\cdot\log n)$ qubits), then an operator is applied to these
that computes the most frequent output and maps it to the real output
qubits. By standard arguments the error probability drops exponentially
in $l$, and so $l=O(1)$ suffices, hence the increase in space is a factor
of $l$ and an additive $O(\log n)$, the increase in time is a factor  of
$l$ and an additive majority computation ($poly(\log n)$ gates) for
each output, the number of queries goes up by a factor of $l$.\end{proof}

Note that we cannot reuse the $O(\log n)$ extra qubits for different outputs, since even if
we try to uncompute the computation on them due to the constant error the result will not be close to
a blank state. The lower bound proof will enforce the space restriction only between slices of the circuit, so
the above construction is still useful. Alternatively we
 could  consider a more general circuit model, in which "fresh" qubits may be added any time and qubits may be "thrown away".
 In such a model the space restriction would refer only to the maximal number of qubits used at the same time.

Due to our output convention now each individual output is correct with
probability $1-\epsilon$ for some arbitrarily small constant $\epsilon$.
But then many outputs must be correct simultaneously with high
probability.

\begin{lemma}\label{lem:simult}
Assume each individual output is correct with probability $1-\epsilon$.
Let $R$ be any set of $l$ outputs. Then with probability
$1-\sqrt{\epsilon}$ at least $(1-\sqrt{\epsilon})l$ of the outputs in
$R$ are correct.
\end{lemma}

\begin{proof}{Proof} If the probability that at least $(1-\sqrt{\epsilon})l$ of
the outputs in $R$ are correct simultaneously is less than
$1-\sqrt{\epsilon}$, then the expected number of correct outputs in $R$
is less than $(1-\sqrt{\epsilon})\cdot
l+\sqrt{\epsilon}\cdot(1-\sqrt{\epsilon})l=(1-\epsilon)l$, which is
impossible due to the linearity of expectation. \end{proof}

Hence with large probability at least a big fraction of the outputs are
correct.

\subsection{Slicing Quantum Circuits}

Consider the following way to slice a given quantum circuit $A$ on $S$ qubits with $T$ queries.
Fix some parameter $\delta$. Slice $A_1$ contains all the gates
from the beginning of the circuit up to the $\delta\sqrt n$-th query gate.
Slice $A_2$ contains the next gates until the $2\delta\sqrt n$-th query
gate and so on. Overall there are $M=\lceil T/(\delta\sqrt n)\rceil$
slices. Note that each slice $A_i$ is a quantum circuit that contains
$\delta\sqrt n$ queries, and that uses $S$ qubits of work space which
are initialized to some state depending on what was computed by
the slices $A_1,A_2,\ldots,A_{i-1}$. Note that for the sorting problem the average number of outputs
in a slice is $n/M$.

In the following we consider the computational power of an individual
slice. We will give an upper bound on the number of outputs a slice can
make.
The set of outputs we consider will be restricted to those which output one of the numbers
smaller than the median, i.e., outputs for the largest $n/2$ numbers will not be considered.
The inputs are of the form $x=x_1,\ldots,x_n$ with all $x_i\in\{1,\ldots,n^2\}$ and $x_i\neq x_j$ for all $i\neq j$.
We can now state our main lemma.

\begin{lemma}[Main]\label{lem:main}
Let $A$ be any quantum algorithm that is initially given some $S$ qubits
in an arbitrary state depending on the classical input $x$, and that afterwards accesses $x$ via
$\delta\sqrt n$ oracle queries only. Assume that $A$
produces $l$ outputs $O_1(x),\ldots,O_{l}(x)$ for the numbers $Min(x,j_1),\ldots,Min(x,j_l)$ with $j_1<\cdots<j_l\le n/2$,
and that for all $x$ and $i\in\{1,\ldots,l\}$ the output $O_i$ is correct
with probability $1-\epsilon$. Then for $\delta=10^{-4}$ it holds that
\begin{eqnarray*}
(1-\sqrt\epsilon)\cdot 2^{-S}\cdot  2^{-l\cdot H(\sqrt{\epsilon})}
\le (0.99)^{(1-\sqrt{\epsilon})l}.\end{eqnarray*}
\end{lemma}

Note that the space bound enters the main lemma only via the amount of
initial information.
The function $H$ is the binary entropy function
$H(p)=-p\log p-(1-p)\log(1-p).$ Let us now deduce
Theorem~\ref{the:lower} from the lemma.

\begin{proof}{Proof of Theorem~\ref{the:lower}} Given is a circuit $A$. First
apply Lemma~\ref{lem:boost} to reduce the error probability to $\epsilon$
for some small enough constant $\epsilon$. Consider any slice $A_i$ of
the obtained circuit and assume that the slice makes some $l$ outputs. Note that the $\log n$ qubits needed for reducing the error
at each output are never reused and hence do not contribute to the initial information.
If $l=cS$ for some large enough constant $c$, then the lemma says
\begin{eqnarray*}
1-\sqrt\epsilon
\le 2^S\cdot 2^{cS\cdot
H(\sqrt{\epsilon})}\cdot(0.99)^{(1-\sqrt{\epsilon})cS}<1/2.\end{eqnarray*}
Contradiction! So the number of outputs in
the slice is at most $O(S)$.

There are $M\le \lceil T/(\delta\sqrt n)\rceil$ slices producing $n/2$
outputs, but the overall number of outputs is at most $\lceil
T/(\delta\sqrt n)\rceil\cdot cS$, so $TS=\Omega(n^{3/2})$.
\end{proof}

\subsection{Proof of the Main Lemma}

The plan of the proof is to show first that in $A$ for each $x$ the expectation over all
subsets containing $l'=(1-\sqrt\epsilon)l$ outputs of the probability that these are
simultaneously correct is at least
$(1-\sqrt\epsilon)\cdot 2^{-S}\cdot 2^{-l\cdot H(\sqrt{\epsilon})}$,
even if the initial information is replaced by a completely mixed state.

Then we show that in any algorithm for any $l'$ output positions
the expectation over inputs $x$ of the probability that the outputs are
simultaneously correct is at most $(0.99)^{l'}$.
These two statements together imply the inequality in the main lemma.
Set $\delta=10^{-4}$.

\begin{lemma}
Suppose an algorithm produces $l$ outputs, and for all inputs $x$ each
output $O_i(x)$ is equal to $(Min(x,j_i),j_i)$ with
probability $1-\epsilon$. Then for all inputs the expectation over all sets
containing $l'$ outputs of the probability that these are simultaneously correct is at least
$(1-\sqrt{\epsilon})\cdot 2^{-H(\sqrt\epsilon)l}$.
\end{lemma}

\begin{proof}{Proof} First apply Lemma~\ref{lem:simult}. This lemma implies
that with probability $1-\sqrt\epsilon$ at least $l'$ outputs are
simultaneously correct.

In other words for all $x$:\[Prob(\exists l'\mbox{ outputs }
O_i(x)=(Min(x,j_i),j_i))\ge 1-\sqrt\epsilon,\] where the probability is over
the measurements. This implies
\[\sum_{1\le o_1 <\ldots <o_{l'}\le l}
Prob(\forall i: O_{o_i}(x)=(Min(x,j_{o_i}),j_{o_i}))\ge 1-\sqrt\epsilon\] and hence
with ${l\choose(1-\sqrt\epsilon)l}\le
2^{lH(\sqrt\epsilon)}$ the lemma follows.\end{proof}

Assume that some $l'$ output gates are supposed to produce the numbers $Min(x,k_1),\ldots,
Min(x,k_{l'})$ for some $k_1<\cdots<k_{l'}\le n/2$ and that these
gates are $O_{k_1}, \ldots,O_{k_{l'}}$ (renumbered for convenience).

Next we get rid of the initial information, at the expense of increasing
the failure probability. We use Lemma~\ref{lem:union}, restated here in
a more specific form.

\begin{lemma}
Suppose there is an algorithm that uses $S$ qubits of initial
information and else makes only queries to the input $x$. Suppose
the algorithm outputs a fixed set $Min(x,k_1),\ldots,
Min(x,k_{l'})$ simultaneously correct with probability $P_x(k_1,\ldots,k_{l'})$.

Then there is an algorithm that uses no initial information, makes the
same number of queries, and has success probability $P_x(k_1,\ldots,k_{l'})/2^S$.
\end{lemma}

So for all $x$ the expectation over all subsets of $l'$ outputs of the success probability is at least
$(1-\sqrt\epsilon)2^{-S}2^{-H(\sqrt\epsilon)l}$. For
a contrasting statement we now consider any fixed set of $l'$ outputs and the expected success
probability over all inputs $x$.
At this point we are left with the following problem. We have a circuit
that is supposed to output $l'$ numbers from the sorted sequence with
some expected success probability $P$. The circuit accesses the input only via
$\delta\sqrt{n}$ queries. We have to show that $P$ is exponentially
small in $l'$.
This is established by the following lemma.

\begin{lemma}\label{lem:query}
For any algorithm that uses $\delta\sqrt n$ queries to inputs
$x$ and tries to output $Min(x,k_1),\ldots, Min(x,k_{l'})$, the expectation (over all $x$) of the success probability is at most
$0.99^{l'}$.
\end{lemma}

Note that this lemma together with the previous two lemmas immediately
implies the main lemma. In the rest of this section we provide its proof.

First note that we can assume that the algorithm never outputs the same number
at two different output gates, since in this case there must be
an error anyway and the respective outputs may be changed arbitrarily.

Our plan is to employ an adversary argument like in Lemma~\ref{lem:condadv}. Therefore we need inputs where we can cheat well.
Let $K=\{k_1,\ldots,k_{l'}\}$. Fix some input $x$. Let $R(x)=\{r_1(x),\ldots,r_{t(x)}(x)\}$ denote a maximal set of positions
from $1,\ldots, n/2$ so that $R(x)\subseteq K$ and $Min(x,r_{i+1}(x))-Min(x,r_i(x))\ge n/8$ for all $i$.
I.e., for each $x$ we single out a set of positions so that the
distance between the elements at these positions in the sorted sequence is almost as large as the average.
\begin{proposition}\label{pro:1}
$t(x)\ge l'/2$ with probability $1-1/2^{l'/2}$.
\end{proposition}

\begin{proof}{Proof}
\begin{eqnarray*}&&Prob(t(x)\le l'/2)\\
&\le&Prob(\mbox{There are } l'/2\,\,\, k\in K:Min(x,k)-Min(x,k-1)<n/8).
\end{eqnarray*}
It is not hard to see that for all $k$
\[Prob(Min(x,k)-Min(x,k-1)<n/8)\le 1/8\] even when conditioned on arbitrarily many events of the form $Min(x,k')-Min(x,k'-1)<n/8$,
since the probability that some intervals are short does rather decrease the probability that another interval is short.
Hence
\begin{eqnarray*}
&&Prob(\mbox{There are } l'/2\,\,\, k\in K:Min(x,k)-Min(x,k-1)<n/8)\\
&\le&(1/8)^{l'/2}\cdot {l'\choose l'/2}<
2^{-l'/2}.\end{eqnarray*}
\end{proof}

Denote by $q_x^i$ the probability
that the outputs $O_{r_1(x)},O_{r_2(x)},\ldots,O_{r_i(x)}$ are correct on $x$. Also denote by $p_x^i$ the probability that
$O_{r_i(x)}$ is correct (on $x$) conditioned on the event that the previous outputs are correct.
Then $q_x^{i+1}=q_x^ip_x^{i+1}$. We are interested in bounding $E[q_x^{l'/2}]$. Let us simply neglect the inputs $x$ for which $t(x)<l'/2$
in the following. These contribute at most $1/2^{l'/2}$ to the success probability due to the above proposition. We plan to show
that there is a constant factor gap between $E[q_x^{i}]$ and $E[q_x^{i+1}]$.

The next proposition states the main adversary argument.
\begin{proposition}
Let $x$ and $i$ be given. There is a set
$J_{x,i}\subseteq\{1,\ldots,n\}$ of size $n/2-\delta n$ so that for each $j\in J_{x,i}$ there are
\[Min(x,r_{i+1}(x))-Min(x,r_{i}(x))-1\] inputs $x'(j)$ so that with probability
$q_x^i(p_x^{i+1}-\sqrt{\delta})$ the
outputs $O_{r_1(x' (j))},O_{r_2(x' (j))},$ $\ldots, O_{r_i(x' (j))}$ are correct and $O_{r_{i+1}(x' (j))}$ is incorrect on $x'(j)$.

Furthermore, each such $x'(j)$ is ``generated'' in this way by at most $n^2$ inputs $x$.
\end{proposition}

\begin{proof}{Proof} We want to apply Lemma~\ref{lem:condadv}. Let the condition $F$ of that lemma be
that the outcomes of the measurements for $O_{r_1(x)}$ up to $O_{r_i(x)}$ are correct. The event $E$ is set to be that the
measurement for $O_{r_{i+1}(x)}$ is correct, i.e., equals $Min(x,r_{i+1}(x))$. Clearly $F$
occurs with probability $q_x^i$ on $x$ and $E$ occurs conditionally with probability $p_x^{i+1}$.
Then the lemma tells us that we may switch $(1-\delta)n$ positions in $x$ arbitrarily, and still
get the same measurement results with probability $q_x^i(p_x^{i+1}-\sqrt{\delta})$. To avoid changing the correctness
of previous outputs we only flip those positions containing numbers larger than $Min(x,n/2)$.
Thus we can flip a set $J_{x,i}$ of at least $n/2-\delta n$ positions.

If we change $x_j$ for $j\in J_{x,i}$ so that its new value $a$ is between $Min(x,r_i(x))$ and $Min(x,r_{i+1}(x))-1$,
then $r_1(x)=r_1(x'(j)),\ldots,r_i(x)=r_i(x'(j))$. Furthermore
the following happens.
\begin{itemize}
\item Either $a-Min(x,r_i(x))<n/8$, and then $r_{i+1}(x'(j))=r_{i+1}(x)+1$. In this case with probability
$q_x^i(p_x^{i+1}-\sqrt\delta)$ on $x'(j)$ the first $i$ outputs in $R(x'(j))$ are correct, and gate
$O_{r_{i+1}(x)}$ outputs $b=Min(x,r_{i+1}(x))$. Since the same number is never output twice the output on $O_{r_{i+1}(x'(j))}$
is not equal to $b$
with the same probability, which is an error.
\item
Otherwise $a-Min(x,r_i(x))\ge n/8$ and so $r_{i+1}(x'(j))=r_{i+1}(x)$. In this case with probability
$q_x^i(p_x^{i+1}-\sqrt\delta)$ on $x'(j)$  the first $i$ outputs in $R(x'(j))$ are correct, and the output $O_{r_{i+1}(x'(j))}$
is $Min(x,r_{i+1}(x))\neq a$, so again the first $i$ outputs in
$R(x'(j))$ are correct and the $i+1$st is not.
\end{itemize}
Note that each $x'(j)$ is derived from at most $n^2$ inputs $x$, since
to change $x'(j)$ to $x$ we have to change one position to some other
value. \end{proof}

Now it is clearly true that
\begin{eqnarray*}
&&E[q_x^i]=E[q_x^i\cdot(p_x^{i+1}+1-p_x^{i+1})],
\end{eqnarray*}
and $q_x^i(1-p_x^{i+1})$ is the probability that the first $i$ outputs in $R(x)$ are correct and the $i+1$st is wrong.

If $Min(x,r_{i+1}(x))-Min(x,r_{i}(x))>8n$, we simply use
$q^i_x(1-p_x^{i+1})\ge0$.

Otherwise we estimate
\begin{eqnarray*}
&&q_x^i(1-p_x^{i+1})\\
&\ge&\max_{y:x=y'(j)}q_y^i (p_y^{i+1}-\sqrt{\delta})\\
&\ge&\sum_{y:x=y'(j)}(q_y^i (p_y^{i+1}-\sqrt{\delta}))/n^2
\end{eqnarray*}
due to the above proposition,
where the notation $x=y'(j)$ indicates that $x$ can be derived from $y$ by a change of the
position $j$ with $x_j=Min(x,r_{i+1}(x))$ as described above. Inserting these estimates and rearranging
the terms in the expectation so that the errors are accounted for
together with the inputs the adversary changes (rather than the inputs resulting from the changes) we get
\begin{eqnarray*}
&&E[q_x^i]
\ge E\left[q_x^ip_x^{i+1}
+q_x^i(p_x^{i+1}-\sqrt{\delta}) \cdot (n/2-\delta n)
\cdot \frac{D(x,i)}{n^2}\right],
\end{eqnarray*}
where
$D(x,i)=\min\{Min(x,r_{i+1}(x))- Min(x,r_{i}(x))-1,8n\}$.
Therefore, recalling that \[Min(x,r_{i+1}(x))-Min(x,r_{i}(x))-1)\ge n/8 \]we get
\begin{eqnarray*}
&&E[q_x^i]+ 1/2\cdot\sqrt\delta E[q_x^i\cdot8]
\ge E[q_x^{i+1}]+(1/2-\delta)\cdot E[q_x^{i+1}\cdot 1/8].
\end{eqnarray*}
Consequentially $E[q_x^i]\cdot .98>E[q_x^{i+1}]$,
and this holds for $i=1,\ldots, l'/2$ (neglecting at most a $1/2^{-l'/2}$ fraction of all inputs), so we have
$E[q_x^{l'/2}]< 2^{-l'/2}+.98^{l'/2}$. Hence the expected probability of getting $O_1,\ldots,O_{l'}$ correct is at most
$.99^{l'}$ for large enough $l'$.

\section{Conclusions and Open Problems}
We have shown that quantum computers, though in general not much faster
for the important task of sorting, outperform classical computers
significantly in space bounded sorting. This setting is motivated by
applications with distributed data too large to fit into the working
memory of a single computer. Furthermore understanding the complexity of
such a basic problem is important in itself. We have seen that for
sorting a tradeoff result of the form $T^2S\le \widetilde{O}(n^3)$
exists, as opposed to the classical tradeoff $TS=\Theta(n^2)$.
Exploring the question whether the upper bound on quantum sorting is tight we have
proved a lower bound of $TS=\Omega(n^{3/2})$, showing that for small
space bounds the algorithm is not too far from optimal. This lower bound actually holds in the average case sense,
i.e., for random sets of $n$ numbers from a $n^2$ range.

Our result can easily be adapted to give the bound
$ST=\Omega(n^2)$ for
classical sorting in the situation that all input positions are known to
hold mutually distinct numbers, by considering circuit slices of length
$\delta n$, and using simple adaptations of Lemma~\ref{lem:union} and Fact~\ref{fac:qm} to the classical case.
\begin{corollary}
Any classical sorting algorithm with time $T$ queries and space $S$ that sorts numbers under the condition that
the input contains mutually distinct elements only, needs $TS=\Omega(n^2)$.\end{corollary}
The best previous bound for sorting under the promise that all numbers are mutually distinct
is $TS=\Omega(n^2\cdot\log\log n/\log n)$ given in \cite{RS82}.

The most important open problem
of this paper is, whether the lower bound can be improved
to (almost) match the upper bound, which is what we conjecture. To
prove such a result it seems one should show that circuit slices
containing $\sqrt{nl}$ queries cannot produce more than $O(l)$ outputs. To do so
it would suffice to show that any quantum algorithm with $\delta\sqrt {nl}$ queries
that tries to compute $l$ elements of the sorted sequence
succeeds only with probability $2^{-\Theta(l)}$.

It is also  of interest what the time-space complexity of sorting is if we allow no
errors. In this situation approaches based on Grover search fail and it
is well possible that the same tradeoff as classically holds.

Consider the element distinctness problem, i.e., deciding whether $n$
given numbers are all pairwise different. It is conjectured that for
classical computers the element distinctness problem has about the same
complexity as sorting, and is thus a decision problem capturing the
difficulty of sorting. Buhrman et al.~describe in \cite{BDH+01} a
quantum algorithm that runs in sublinear time, and a slight variation of
their algorithm achieves a tradeoff of $T^2S=\widetilde{O}(n^2)$ for
deciding element distinctness. Hence in the quantum case element
distinctness is
strictly easier than sorting due to the lower bound for sorting given in
this paper, consider e.g.~the case that $S\le poly(\log n)$. Can a
matching lower bound be shown for element distinctness? Is element
distinctness really as hard as sorting classically? Note that strong
classical tradeoffs are known for element distinctness if only a comparison oracle
is used \cite{BFM+87,Y88}, but the best tradeoff known for general
models is $T=\Omega(n\log(n/S))$, given in \cite{BSSV00}, no better product tradeoff than $ST=\Omega(n\log^2n)$.
A quantum query lower bound of $\Omega(n^{2/3})$ has recently been shown by Shi
\cite{S02}.

Another open problem concerns the query problem we have analyzed. We have shown that finding $l$ elements of the sorted sequence within
$\delta\sqrt n$ queries is possible with exponentially small success probability only. More generally, suppose one is given $l$
instances of a query problem, and is allowed to do a number $q$ of queries that is known to give the result for one
instance with probability $p<1$ only, on average over all inputs. How large is the success probability of computing
correctly on $l$ instances? In other words, does a direct product
result hold for quantum black-box algorithms? Such theorems for classical query algorithms are given in \cite{IRW94,NSR99}.

\section*{Acknowledgments}
The author wishes to thank Peter H\o yer, Alexander Raz\-borov, Avi Wigderson,
Ronald de Wolf, and the anonymous referees for helpful comments and discussions.


\begin{thebibliography}{10}
\vspace{.3cm}
\bibitem{AKN98}
D.~Aharonov, A.~Kitaev, and N.~Nisan. Quantum circuits with mixed
states. {\em 30th {ACM} Symposium on Theory of Computing}, pp.~20--30,
1998.
Also: {\tt quant-ph/9806029}.

\bibitem{A00} A.~Ambainis.
Quantum lower bounds by quantum arguments. {\em 32nd {ACM} Symposium on Theory of Computing}, pp.~636--643, 2000.
Also: {\tt quant-ph/0002066}.

\bibitem{B91} P.~Beame. A general sequential time-space tradeoff for
finding unique elements. {\em SIAM Journal on Computing,} vol.20,
pp.270--277, 1991.

\bibitem{BSSV00} P.~Beame, M.~Saks, X.~Sun, E.~Vee. Super-linear
time-space tradeoff lower bounds for randomized computation. {\em 41st
IEEE Symposium on Foundations of Computer Science}, pp.169--179, 2000.

\bibitem{BBBV97} C.H.~Bennett, E.~Bernstein, G.~Brassard, and
  U.~Vazirani. Strengths and weaknesses of quantum computing. {\it
  SIAM Journal on Computing}, vol.~26, pp.~1510--1523, 1997. Also: {\tt
  quant-ph/9701001.}

\bibitem{BBC+93} C.H.~Bennett, G.~Brassard, C.~Crepeau, R.~Josza,
A.~Peres, W.~Wooters. Teleporting an Unknown Quantum State via Dual
Classical and Einstein-Podolsky-Rosen Channels. {\em Phys.~Rev.~Lett.},
vol.70, pp.1895--1899, 1993.

\bibitem{BC82} A.~Borodin, S.~Cook. A time-space tradeoff for sorting on
a general sequential model of computation. {\em SIAM Journal on
Computing,} vol.11, pp.287--297, 1982.

\bibitem{BFM+87} A.~Borodin, F.~Fich, F.~Meyer aud der Heide, E.~Upfal, A.~Wigderson.
A time-space tradeoff for element distinctness. {\em SIAM Journal on Computing,} vol.16, pp.97--99, 1987.

\bibitem{BDH+01} H.~Buhrman, C.~D\"urr, M.~Heiligman, P.~H\o yer,
F.~Magniez, M.~Santha, R.~de Wolf. Quantum Algorithms for Element
Distinctness. {\em IEEE Conference on Computational Complexity},
pp.120--130, 2001.

\bibitem{BW02} H.~Buhrman, R.~de Wolf. Complexity Measures and Decision
Tree Complexity: A Survey. To appear in {\em Theoretical Computer
Science}, 2002.

\bibitem{CLRS01} T.H.~Cormen, C.E.~Leiserson, R.L.~Rivest, C.~Stein.
Introduction to Algorithms. MIT Press, 2001.

\bibitem{DH96} C.~D\"urr, P.~H\o yer. A quantum algorithm for finding
the minimum. {\tt quant-ph/9607014}, 1996.

\bibitem{G96} L.K.~Grover. A fast quantum
  mechanical algorithm for database search. {\it 28th ACM Symposium on
    Theory of Computing}, pp.~212-219, 1996. Also: {\tt
quant-ph/9605043.}

\bibitem{HNS01} P.~H\o yer, J.~Neerbek, Y.~Shi. Quantum complexities of
ordered searching, sorting, and element distinctness.
{\em 28th International Colloquium on Automata, Languages, and Programming}, pp.62--73,
2001. Also: {\tt quant-ph/0102078.}

\bibitem{IRW94} R.~Impagliazzo, R.~Raz, A.~Wigderson.  A Direct Product Theorem.
{\em IEEE Conference on Structures in Complexity Theorry}, pp.88--96, 1994.

\bibitem{NC00} M.A.~Nielsen and I.L.~Chuang. Quantum Computation and
  Quantum Information. Cambridge University Press, 2000.

\bibitem{NSR99} N.~Nisan, M.~Saks, S.~Rudich. Products and Help Bits in Decision trees.
{\em SIAM Journal on Computing,} vol.28, pp.1035--1050, 1999.

\bibitem{PR98} J.~Pagter, T.~Rauhe. Optimal Time-Space Trade-Offs for
Sorting.  {\em 39th IEEE Symposium on Foundations of Computer Science},
pp.264--268, 1998.

\bibitem{RS82} S.~Reisch, G.~Schnitger. Three Applications of Kolmogorov-Complexity.
{\em 23rd IEEE Symposium on Foundations of Computer Science},
pp.45--52, 1982.

\bibitem{S02}  Y.~Shi. Quantum Lower Bounds for the Collision and the
Element Distinctness Problems.  {\em 43rd IEEE Symposium on Foundations of Computer
Science}, pp.513--519, 2002.

\bibitem{Y88} A.C.C.~Yao. Near optimal time-space tradeoffs for element
distinctness. {\em 29th IEEE Symposium on Foundations of Computer
Science}, pp.91--97, 1988.
\end{thebibliography}
\end{document}